\documentclass{iau} 
\usepackage{graphicx}
\usepackage{url}

\title[The metallicity dependence of WR winds] 
{The metallicity dependence of WR winds}

\author[R. Hainich, T. Shenar, A. Sander, W.-R. Hamann \& H. Todt]   
       {R. Hainich$^1$,
        T. Shenar$^1$,
        A. Sander$^1$,
        W.-R. Hamann$^1$,
        \and H. Todt$^1$
       }

\affiliation{$^1$Institut f\"ur Physik und Astronomie,
              Universit\"at Potsdam, \\
              Karl-Liebknecht-Str. 24/25,
              D-14476 Potsdam, Germany \\
              email: {\tt rhainich@astro.physik.uni-potsdam.de}
            }

\pubyear{2017}
\volume{329}  
\jname{The lives and death-throes of massive stars}
\editors{JJ Eldridge, B.D. Editor \& C.E. Editor, eds.}
\begin{document}

\maketitle

\begin{abstract}
Wolf-Rayet (WR) stars are the most advanced stage in the evolution of the most massive stars. The strong feedback provided by these objects and their subsequent supernova (SN) explosions are decisive for a variety of astrophysical topics such as the cosmic matter cycle. Consequently, understanding the properties of WR stars and their evolution is indispensable. A crucial but still not well known quantity determining the evolution of WR stars is their mass-loss rate. Since the mass loss is predicted to increase with metallicity, the feedback provided by these objects and their spectral appearance are expected to be a function of the metal content of their host galaxy. This has severe implications for the role of massive stars in general and the exploration of low metallicity environments in particular. Hitherto, the metallicity dependence of WR star winds was not well studied. In this contribution, we review the results from our comprehensive spectral analyses of WR stars in environments of different metallicities, ranging from slightly super-solar to SMC-like metallicities. 
Based on these studies, we derived empirical relations for the dependence of the WN mass-loss rates on the metallicity and iron abundance, respectively.

\keywords{Wolf-Rayet stars, Magellanic Clouds, stellar atmospheres, stellar winds, mass-loss, metallicity}
\end{abstract}

\firstsection 
\section{Introduction}

The line-driven winds of hot stars are expected to be metallicity ($Z$) dependent, since they are predominately driven by the interaction of photons with millions of iron lines in the extreme UV. While this mechanism has been studied both empirically and theoretically for OB-type stars, a thorough investigation for Wolf-Rayet (WR) stars was pending for a long time. In the WR phase, strong and powerful stellar winds are capable of shedding a significant amount of the stellar mass, determining 
the conditions under which these objects might or might not explode as supernovae (SNe, see e.g.\ \cite[Heger et al. 2003, Dessart et al. 2011, Groh et al. 2013]{Heger2003,Dessart2011,Groh2013}).
Therefore, it is urgent to understand how the properties of WR winds change as a function of $Z$. This is all the more important in the light of the recent gravitational wave discoveries (\cite[Abbot et al. 2016a, b]{Abbott2016a,Abbott2016b}). First empirical studies of the metallicity dependence of WR stars were performed by \cite{Crowther2006} and \cite{Nugis2007}, while theoretical approaches were presented by \cite{Graefener2008} and \cite{Vink2005}.

In a series of paper, we presented analyses of an unprecedented number of WR stars of the nitrogen sequence (WN stars), covering metallicities from solar down to the Small Magellanic Cloud (SMC). In the SMC, we investigated the complete population, consisting of seven potentially single stars \cite[(Hainich et al. 2015)]{Hainich2015} and five binary systems \cite[(Shenar et al. 2016)]{Shenar2016}. The Large Magellanic Cloud (LMC) sample encompasses almost all putative single stars \cite[(Hainich et al. 2014)]{Hainich2014}, while the analysis of the binary systems is currently underway. 
Solar metallicity is covered by the analyses of Galactic WN stars  \cite[(Hamann et al. 2006, Martins et al. 2008, Liermann et al. 2010, Oskinova et al. 2013)]{Hamann2006,Martins2008,Liermann2010,Oskinova2013}.
These comprehensive data sets allow us to investigate the relation between the mass-loss rate and the metallicity for WN stars, as presented in this contribution. 

\section{Stellar atmosphere models}

The spectral analyses employed in this work were mostly performed with the Potsdam Wolf-Rayet (PoWR) model atmospheres. The PoWR code is a state-of-the-art stellar atmosphere code (see e.g. Sander 2017, these proceedings) that, among other things, allow for deviations from LTE, complex model atoms, iron line-blanketing, wind inhomogeneities, and a consistent treatment of the quasi-hydrostatic domain. The statistical equations are solved simultaneously with the radiative transfer that is calculated in the co-moving frame, while energy conservation is ensured. 

\begin{figure}[hbtp]
\begin{center}
 \includegraphics[width=\textwidth]{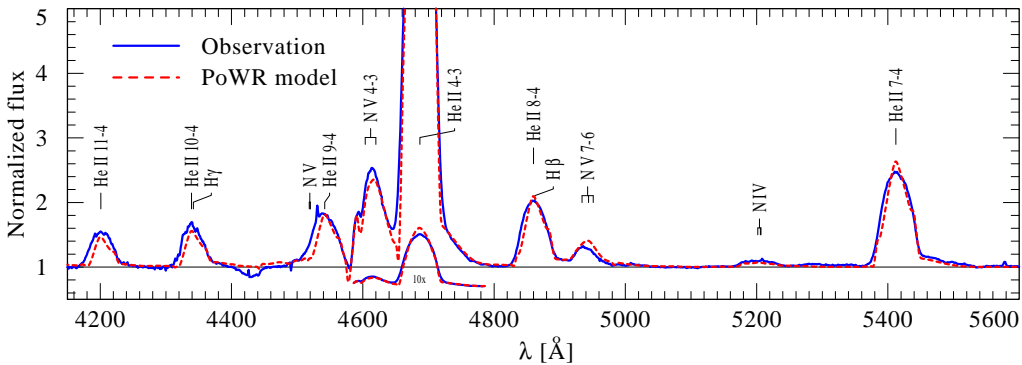} 
 \caption{Spectral fit for WR\,18, an early-type Galactic WN star. The observation is depicted as a solid blue line, while the PoWR model is shown as a red dashed line.}
   \label{fig:fit}
\end{center}
\end{figure}

The analysis of complete WN populations is only feasible by means of large model grids that we have computed for a variety of metallicities. These model grids, as well as additional grids for WC and OB-type stars, are publicly available via our PoWR website\footnote{\url{www.astro.physik.uni-potsdam.de/PoWR}} \cite[(Todt et al. 2015)]{Todt2015}. For a detailed description of grid based analyses, we refer to \cite{Sander2012} and \cite{Hainich2014}. We note that additional models with adjusted abundances and terminal velocities were calculated for a subset of the stars discussed in this work. A typical fit of the normalized optical line spectrum is presented in Fig.\,\ref{fig:fit}. 

\section{The mass-loss rate as a function of metallicity}

A comparison between the spectra of typical WN stars from different metallcity regimes clearly reveals that the emission line strengths are declining with decreasing metallicity, indicating a profound dependency of the wind strength on $Z$. To derive an empirical relation between the mass-loss rate and the metallicity, we utilized the ``modified wind momentum'' \cite[(Kudritzki et al. 1995, Puls et al. 1996, Kudritzki et al. 1999)]{Kudritzki1995,Puls1996,Kudritzki1999}, which is defined as $D_\mathrm{mom} = \dot{M} v_\infty R_*^{1/2}$. Here, we neglect a potential $Z$ dependency of the terminal wind velocity $v_\infty$, since this dependence is weak, if present at all \cite[(Niedzielski et al. 2004)]{Niedzielski2004}. Consequently, $D_\mathrm{mom}$ is expected to show the same metallicity dependence as the mass-loss rate.

\begin{figure}[btp]
    \begin{center}
       \includegraphics[width=0.56\textwidth]{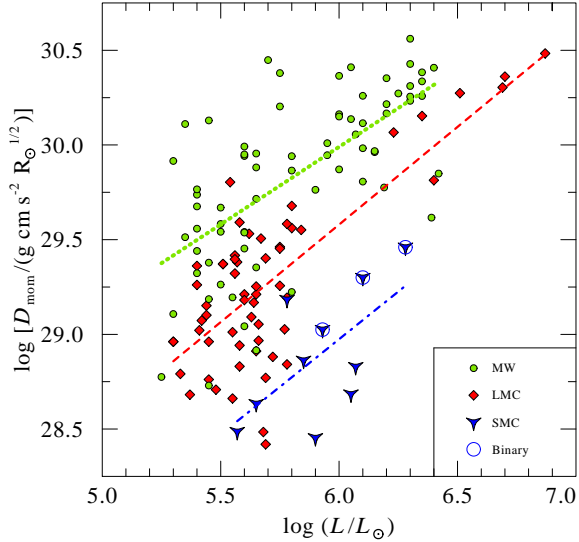} 
       \caption{Modified wind momentum as a function of the luminosity for the galactic WN stars (green symbols), LMC WN stars (red symbols), and SMC WN stars (blue symbols). The lines are the WLRs fitted to the three populations.}
       \label{fig:wmlr}
    \end{center}
\end{figure}

A tight relation is predicted between $D_\mathrm{mom}$ and the luminosity $L$, the so-called wind-momentum luminosity relation \cite[(WLR,\ Kudritzki et al. 1999)]{Kudritzki1999}. This allows us to implicitly account for the dependence of the wind strength on the luminosity, while utilizing the WLRs of the distinct WN population to determine the $\dot{M}$-$Z$-relation. The modified wind momentum of the WN stars in the MW, LMC, and SMC are plotted in Fig.\,\ref{fig:wmlr} as a function of the luminosity. This figure also depicts the three WLRs fitted to the different WN populations. The coefficients of the respective relations are given in Table\,\ref{tab:wmlr}.

\begin{table}
    \begin{center}
        \caption{Coefficients of the wind-momentum luminosity relations $D_\mathrm{mom} = D_{0} \cdot L^{\alpha}$ fitted to the WN populations in the MW, LMC, and SMC, respectively.}
        \label{tab:wmlr}
        \begin{tabular}{lcc}\hline 
            & $\log D_0$            & $\alpha$  \\ \hline 
            {\bf MW  } & $25.1 \pm 0.7$ & $0.8 \pm 0.1$   \\ 
            {\bf LMC } & $23.4 \pm 0.6$ & $1.0 \pm 0.1$   \\ 
            {\bf SMC } & $22.9 \pm 2.6$ & $1.0 \pm 0.4$   \\ 
            \hline
        \end{tabular}
    \end{center}
\end{table}

\begin{figure}[btp]
    \begin{center}
       \includegraphics[width=0.55\textwidth]{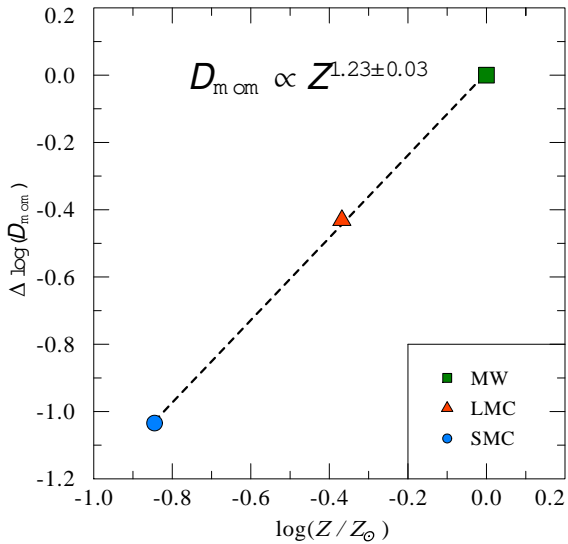} 
       \caption{Offsets between the wind-momentum luminosity relations plotted over the difference between the metallicities of the corresponding galaxies.}
       \label{fig:mdotz}
    \end{center}
\end{figure}

Since the slope of the WLR fitted to the Galactic population is slightly different compared to the other two WLRs, we have to either evaluate these equations at a specific luminosity (e.g.\, $\log L/L_\odot = 5.9$), or assume the same slope for all three WLRs. In fact, both methods give the nearly identical results. 
For Fig.\,\ref{fig:mdotz}, we have evaluated the WLRs (shown in Fig.\,\ref{fig:wmlr}) at a luminosity of $\log (L/L_\odot) = 5.9$. The differences between the WLRs on the $D_\mathrm{mom}$-scale are plotted over the difference in the metallicities of the corresponding host galaxies. A linear fit to these data points reveals the metallicity dependence of $D_\mathrm{mom}$ and, consequently, of the mass-loss rate:
\begin{equation}
\label{eq:mdotz}
\dot{M} \propto Z^{1.2 \pm 0.1}~.
\end{equation}  
We note that the quoted uncertainty only represents the statistical error of the linear regression. 

Since the main contribution to the radiative acceleration of the line driven winds of massive stars is provided by interaction of the stellar radiation field with the multitude of iron lines in the extreme UV, we alternatively investigated the dependence of the wind momentum specially on the iron abundance $X_\mathrm{Fe}$. 
Due to the lack of high resolution and high signal-to-noise UV data for most of the MW WN stars, individual iron abundances are not available for most of these objects. Therefore, we used the iron abundance measured for the Galactic stellar population as a proxy for the iron abundance of the individual WN stars.
In the outer part of the our Galaxy, the iron abundance shows a well established gradient \cite[(Hayden et al. 2015)]{Hayden2015}, while it seems to be rather constant and slightly supersolar in the inner part \cite[(Cunha et al. 2007, Najarro et al. 2009, Ryde \& Schultheis 2015)]{Cunha2007,Najarro2009,Ryde2015}. 
Consequently for stars with a galactocentric distance of less than 6\,kpc, we assume an iron abundance of $X_\mathrm{Fe} = 0.0014$ (mass fraction), while for stars outwards of this radius, it is reduced according to the gradient in the Galactic iron abundance.

\begin{figure}[btp]
    \begin{center}
        \includegraphics[width=0.56\textwidth]{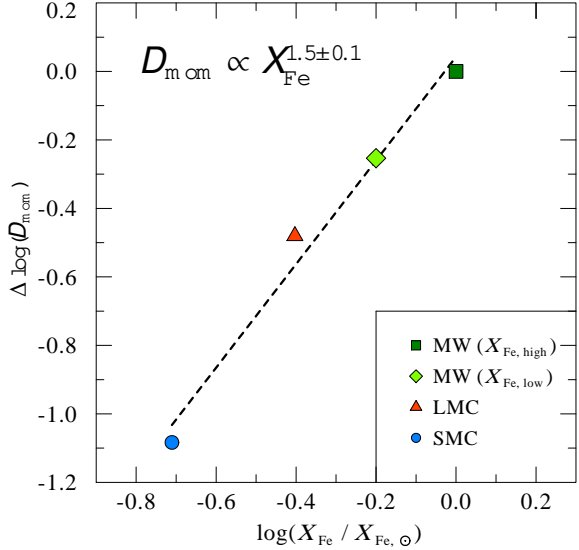} 
        \caption{Differences between the wind-momentum luminosity relations plotted over the differences between the iron abundance of the massive stars in the corresponding galaxies.}
        \label{fig:wmlrfe}
    \end{center}
\end{figure}

Based on these assignments, the Galactic WN population can be separated in two groups, one with a high and one with a relatively low iron abundance. As for the metallicity, the LMC, SMC, and the two Galactic samples can be evaluated by means of their WLRs. In Fig.\,\ref{fig:wmlrfe}, the differences between the WLRs on the $D_\mathrm{mom}$-scale are plotted over the iron abundances of the respective populations.
Fitting these data points with a linear regression provides the following relation between the mass-loss rate of WN stars and their iron abundance:
\begin{equation}
\label{eq:mdotfe}
\dot{M} \propto X_{\mathrm{Fe}}^{1.5 \pm 0.1}~.
\end{equation}  

\section{Implications}

Based on comprehensive spectral analyses of the WN stars in the MW, LMC, and SMC, we determine the dependence of the mass-loss rate on the environments' total metallicity and, alternatively, specifically the iron abundance. The latter proves to be steeper than the former, pointing to the importance of iron for the line driving. However, the uncertainty of this relation is large. 

In comparison to what was found for OB-type stars \cite[(see e.g.\ Vink et al. 2007, Mokiem et al. 2007)]{Vink2001,Mokiem2007}, the metallicity dependence of the WN stars is found to be significantly steeper: $\dot{M}_\mathrm{OB} \propto Z^{0.8}$ vs. $\dot{M}_\mathrm{WN} \propto Z^{1.2}$. This probably reflects the growing importance of multiple scattering for the wind driving with higher metallicity.


%
%
%

\end{document}